\documentstyle[12pt]{article}
\begin{document}

\baselineskip=24pt plus 2pt
\vskip -5cm
\hfill\hbox{NCKU-HEP/96-04}

\vskip 1cm
 
\begin{center}
{\large \bf The $U_L(3)\times U_R(3)$ Extended Nambu--Jona-Lasinio\\
Model in Differential Regularization}\\
\vspace{20mm}
Yaw-Hwang Chen, Su-Long Nyeo \\
and\\
Yeou-Wei Yang\\
\vspace{5mm}

Department of Physics, National Cheng Kung University\\
Tainan, Taiwan 701, R.O.C \\

\vspace{10mm}
\end{center}
\begin{center}
{\bf ABSTRACT}
\end{center}
We employ the method of differential regularization to calculate 
explicitly the one-loop effective action of a bosonized 
$U_L(3)\times U_R(3)$ extended Nambu--Jona-Lasinio model
consisting of scalar, pseudoscalar, vector and axial vector fields.

\newpage
\noindent
{\large \bf 1. Introduction}

Quantum chromodynamics (QCD) is a theory of strong interactions [1]
and is asymptotically free such that the forces between quarks become weak 
for small quark-quark separation or, equivalently, large momentum transfer.  
The same self-interactions of gluons that give rise to asymptotic freedom 
lead to a strong quark-quark interaction for medium and small energies, 
and thus the low energy hadron physics cannot be handled by perturbative QCD.
At low energies, chiral perturbation theory [2] is a good method for 
perturbative calculations.  We note that a disadvantage of chiral 
perturbation theory is that as soon as we go beyond the lowest order, 
the number of free parameters 
increases very rapidly, thus making calculations beyond the lowest few 
orders rather impractical.  We would thus like to obtain these free 
parameters directly from QCD.  This is rather difficult to do so far, and
there is a need for some models that interpolate between 
QCD and chiral perturbation theory.

Among many models, the Nambu--Jona-Lasinio (NJL) model [3,4] seems 
to be the simplest pure quark theory, which yields dynamical symmetry 
breaking and hence a nonvanishing value for the quark condensate.  
But to make the perturbation series nondivergent, we need to introduce 
an ultraviolet cutoff as a regularization.  Many regularization methods 
have been employed [5] and it was found that the one-loop effective 
potentials in various methods can vary substantially with respect to 
the change of the renormalization scale.  In particular, one should 
know that not all regularization methods are suitable to
NJL models.  Moreover, higher loop corrections are also needed to
reduce the sensitivity of the renormalization scheme dependence
of perturbative results.

In this paper, we shall use a space-time regularization method known
as differential regularization (DR) [6,7] to calculate the one-loop effective
action of an extended NJL (ENJL) model.  This method has been shown to 
be useful for studying the quantum corrections in chiral theories.  By using 
DR method, we can determine the effective action systematically and 
unambiguously.  We organize the paper as follows.  In section 2, we 
introduce the ENJL model and the calculational procedure of DR method.  
In section 3, we calculate explicitly the one-loop effective action of 
the ENJL model.

\vskip 1cm
\noindent
{\large \bf 2. The Extended NJL Model and Differential Regularization}

We consider the $U_L(3)\times U_R(3)$ extended NJL Lagrangian for 
quark field [8,9]:
\begin{eqnarray}
{\cal L}_{ENJL}& = &\bar{q}(i\gamma^\mu\partial_\mu - m_0)q\nonumber\\
        &   &+ 2G_1\sum_{i=0}^{N_f^2 -1}\left[(\bar{q}
            {\lambda^i\over 2} q)^2
            + (\bar{q}{\lambda^i\over 2}i\gamma_5 q)^2 \right]\nonumber\\
        &   &- 2G_2\sum_{i=0}^{N_f^2 -1}\left[(\bar{q}{\lambda^i\over2}
            \gamma_\mu q)^2
            + (\bar{q}{\lambda^i\over2}i\gamma_5\gamma_\mu q)^2\right]\,,
\end{eqnarray}
where $G_1$ and $G_2$ are the four-fermion coupling constants, $m_0$ is the
quark mass matrix and $\lambda^i$ are Gell-Mann matrices.  

The quantized theory can be written in terms of a generating 
functional  which, in the absence of external sources, reads
\begin{equation}
Z_{ENJL} = \int D\bar{q}Dq\exp\left[i\int 
d^4x{\cal L}_{ENJL}(x)\right]\,.
\end{equation}
In order to use the model to describe the low energy properties with the
manifest low energy modes, we bosonize the model with the auxiliary
fields $\Phi=\Phi_\alpha\Lambda_\alpha$ introduced via the identity
\begin{eqnarray}
& & \exp\left(-{i\over2}\int\bar{q}\Lambda_\alpha qQ^{\alpha\beta}\bar{q}
\Lambda_\beta q\right)\nonumber\\
& & {\hskip 2cm}= \int D\Phi\exp\left(-{i\over2}\int\Phi_\alpha
(Q^{-1})^{\alpha\beta}\Phi_\beta - i\int\Phi_\alpha\bar{q}
\Lambda_\alpha q\right)\,.
\end{eqnarray}
It contains therefore (in the case of three flavors) nonets of scalar,
pseudoscalar, vector and axial vector meson fields:
\begin{equation}
\Lambda_\alpha = {\lambda^i\over 2}\otimes\Gamma_a,~~i=0,\cdots,N_f^2-1
,~~\Gamma_a\in\{1,~~i\gamma_5,~~i\gamma_\mu,~~i\gamma_\mu\gamma_5\}\,,
\end{equation}
\begin{equation}
Q^{\alpha\beta} = \cases{&${\hskip -0.5cm}4G_1
              \delta^{\alpha\beta}~~{\rm for}~~\Gamma_a\in\{1,~~i\gamma_5\}$\cr
                        &${\hskip -0.5cm}4G_2
              \delta^{\alpha\beta}~~{\rm for}~~\Gamma_a
                        \in\{\gamma_\mu,~~\gamma_\mu\gamma_5\}$}\,.
\end{equation}
Hence we have
\begin{equation}
{\cal Z}_{ENJL} = \int D\Phi\exp\left(-{i\over2}\int\Phi Q^{-1}\Phi\right)
                Z_F\left[\Phi\right]\,,
\end{equation}
where
\begin{equation}
Z_F\left[\Phi\right] 
     = \int Dq D\bar{q}\exp\left(-i\int\bar{q}(i\gamma^\mu\partial_\mu
       - m_0 - \Phi)q\right)\,.
\end{equation}
By shifting $\Phi\rightarrow\Phi - m_0$, we have
\begin{equation}
{\cal Z}_{ENJL} = \int D\Phi e^{-{i\over2}\int d^4x(\Phi - m_0)Q^{-1}
           (\Phi - m_0)}
           \int Dq D\bar{q}e^{-i\int d^4x\bar{q}(i\gamma^\mu
           \partial_\mu - \Phi)q}\,.
\end{equation}
The auxiliary field $\Phi$ can be expressed in terms of the scalar, 
pseudoscalar, vector and axial vector fields as
\begin{equation}
\Phi = g_SS + ig_P\gamma_5 P - ig_V\gamma^\mu V_\mu 
- ig_A\gamma^\mu A_\mu\gamma_5\,.
\end{equation}
Then we can write
\begin{eqnarray}
{1\over2}(\Phi - m_0)Q^{-1}(\Phi - m_0) &=&
{1\over2G_1}Tr((S-m_0)^2 + P^2)\nonumber\\
& &+ {1\over2G_2}Tr(V_\mu V^\mu + A_\mu A^\mu)\,,
\end{eqnarray}
where $S = \sum^8_{i=0}{\lambda^i\over2}S^i$,
$P = \sum^8_{i=0}{\lambda^i\over2}P^i$,
$V_\mu = {\lambda^0\over2}\omega_\mu +
\sum^8_{i=1}{\lambda^i\over2}\rho^i_\mu$, 
and $A_\mu = {\lambda^0\over2}f_\mu +
\sum^8_{i=1}{\lambda^i\over2}a^i_\mu$.

By including the electroweak interaction, we have the generating functional
\begin{eqnarray}
{\cal Z}_{ENJL}& = &\int DW DB D\Phi 
           e^{-\int d^4x{i\over2}(\Phi - m_0)Q^{-1}
           (\Phi - m_0) - {1\over4}F_{\mu\nu}F^{\mu\nu}
           - {1\over4}G_{\mu\nu}G^{\mu\nu}}\nonumber\\
     &   & \times\int Dq D\bar{q}e^{-i\int d^4x\bar{q}(i\gamma^\mu
           \partial_\mu - \Phi)q}\,,
\end{eqnarray}
where
\begin{eqnarray}
F_{\mu\nu} & = & \partial_\mu B_\nu - \partial_\nu
                 B_\mu\,,\nonumber\\
G_{\mu\nu} & = & \partial_\mu W_\nu - \partial_\nu W_\mu
                 +g_W\left[W_\mu, W_\nu\right]\,.
\end{eqnarray}
Then the auxiliary field $\Phi$ is expressed as the scalar, 
pseudoscalar, vector, axial vector fields and gauge bosons,
\begin{equation}
\Phi = g_SS + ig_P\gamma_5P - ig_V\gamma^\mu V_\mu - ig_A\gamma^\mu A_\mu
       \gamma_5 -ig_W\gamma^\mu W_\mu - ig_B\gamma^\mu B_\mu\,,
\end{equation}
where $W$ and $B$ are the electroweak gauge bosons.
The fields $S$ and $P$ can be represented chirally as 
\begin{equation}
S+i\gamma_5 P = P_R M + P_L M^\dagger\,,
\end{equation}
where $P_R = {1\over2}(1 + \gamma_5)$ and 
$P_L = {1\over2}(1 - \gamma_5)$.  We can parameterize $M = S + iP$
as $M = U\Sigma U$, with $\Sigma$ being the scalar $\sigma$ fields, and
$U$ the coset $U_L(3)\times U_R(3)/U_V(3)$:
\begin{eqnarray}
U & = & \exp\left\{i\sqrt{2}{\Phi_8 + \Phi_1\over f_0}\right\}\,,\\
\Phi_8 & = & \sum^8_{i=1} {\lambda^i\over\sqrt{2}}\phi^i\nonumber\\
         & = & \left(\begin{array}{ccc}
                {1\over\sqrt{2}}\pi^0 + {1\over\sqrt{6}}\eta_8
             & \pi^+ & k^+\\
               \pi^- 
             & -{1\over\sqrt{2}}\pi^0 + {1\over\sqrt{6}}\eta_8
             & k^0 \\
               k^- & \bar{k^0}
             & -{2\over\sqrt{6}}\eta_8
              \end{array}\right)\,,\\
\Phi_1 & = & {1\over\sqrt{3}}\eta_1 I\,.
\end{eqnarray}

In the low-energy approximation, we couple the scalar, pseudoscalar, vector 
and axial vector fields to external sources.  The quark fields are not 
coupled to any external sources and can be integrated over to give a 
functional determinant, whose evaluation requires a regularization. 
The calculation can be carried out by using the Feynman rules for the quark 
fields, such that the one-loop quantum effects are considered by treating 
quarks as internal lines in the Feynman diagrams.

In this model, it is necessary to determine its vacuum by calculating the
effective potential or action to at least one-loop order.  We shall employ 
differential regularization and carry out the calculation in Euclidean space, 
so that the functional integral reads
\begin{equation}
Z_{ENJL}' = \int D\Phi
\exp\left[-{\cal S}_{eff}(\Phi)\right]\,,
\end{equation}
where the effective action in the one-loop fermion approximation is
\begin{eqnarray}
{\cal S}_{eff} &=& - \ln{\rm Det}[-i\gamma^\mu\partial_\mu +
              \Phi]\nonumber\\
               & & + {1\over2}\int d^4x
              (\Phi - m_0)Q^{-1}(\Phi - m_0)\,.
\end{eqnarray}

The functional determinant can be calculated using the Feynman rules;
in particular, the massless quark propagator is given by
\begin{eqnarray}
\langle {q_a^i(x)\bar{q}_b^j(0)}\rangle
&\equiv& \Delta_{ab}^{ij}(x)\nonumber\\
&=& - {i{\delta^{ij}\delta_{ab}}\over{2\pi^2}}{\gamma_\mu
         x^\mu\over x^4}\,,
\end{eqnarray}
with $i,j$ and $a,b$ being the isospin and color indices, respectively.

In perturbative calculations, we encounter highly singular terms of the form
\begin{equation}
{1\over(x^2)^n}\ln^m(\mu^2x^2),~~n \geq 2\,, m\geq 0\,,
\end{equation}
where $\mu$ is a mass parameter in the problem.  The essential idea 
of differential regularization is to define these highly singular 
terms by
\begin{equation}
{1\over(x^2)^n}\ln^m(\mu^2x^2)
\equiv \underbrace{\Box\Box\ldots\Box}_{n - 1}G(x^2)\,,x^2\neq 0\,,
\end{equation}
where $G(x^2)$ is a to-be-determined function that has a well-defined 
Fourier transform and can depend on $2(n-1)$ integration constants, which 
play the role of a subtraction scale.  In this paper, we encounter only the 
following two forms:
\begin{equation}
{1\over(x^2)^2} \rightarrow - {1\over4}\Box{\ln x^2\mu^2\over x^2}\,,
x^2\neq 0\,,
\end{equation}
\begin{equation}
{1\over(x^2)^3} \rightarrow - {1\over32}\Box\Box{\ln x^2\mu^2\over x^2}\,,
x^2\neq 0\,,
\end{equation}
where the mass parameter $\mu$ is an integration constant.  Note that we 
have omitted other irrelevant integration constants for $x^2\neq 0$.  
This regularization method has been used in $\phi^4$ theory, QCD [6] and 
Nambu--Jona-Lasinio model [7] and can reproduce the well-known results 
obtained by other methods.  The advantage of DR method is that loop 
corrections in a chiral theory can be calculated unambiguously.  We 
note also that different methods can lead to different results for 
the one-loop effective potential, which depends strongly on 
renormalization scheme.  Higher loop corrections can reduce the 
sensitivity of scheme dependence, but of course, the use of a regularization 
method requires special attention.

\vskip 1cm
\noindent
{\large \bf 3. The One-Loop Effective Action}

To obtain the one-loop effective action of the ENJL model, we need to 
evaluate the one-loop bilinear terms in the scalar, pseudoscalar, vector 
and axial vector fields, and their interaction terms. For 
illustrative purposes, we evaluate some of them below.  The bilinear
term in the scalar field with an internal quark loop is easily calculated 
and reads
\begin{eqnarray}
\Pi_S(x, y) & = & - {g^2_S\over2}\int d^4x d^4y
            {\rm Tr}\left[\Delta(x-y)\Delta (y-x)
            S(x)S(y)\right]\nonumber\\
            & = & - {3g^2_S\over2\pi^4} \int d^4x d^4y
            {\rm Tr}\left[S(x)S(y)\right]
            {1\over(x-y)^6}\nonumber\\
            &\buildrel sing. \over = & 
            {{3g^2_S}\over{128\pi^4}} \int d^4x d^4y
            \left[{3\over2}S^0(x)S^0(y) + \sum^8_{i=1}
            S^i(x)S^i(y)\right]\nonumber\\
            &   & \times\Box\Box{\ln(x-y)^2\mu^2\over(x-y)^2}\,,
            {(x - y)}^2\neq 0\,,
\end{eqnarray}
where $\Box = \Box_{(x-y)}$ and we have used ${\rm Tr}_\gamma I = 4,
{\rm Tr}_cI = 3,$ and ${\rm Tr}_\lambda I = 3$ for the spinor, color,
and octet degrees of freedom, respectively.  The logarithmic dependence 
of the scalar term reads
\begin{equation}
{d\Pi_S\over d\ln\mu^2} 
         = - {3g^2_S\over32\pi^2}\int d^4x
          \left[{3\over2}S^0(x)\Box S^0(x) + \sum^8_{i=1}
          S^i(x)\Box S^i(x)\right]\,.
\end{equation}

The bilinear term in the pseudoscalar field is
\begin{eqnarray}
\Pi_P(x,y) & = & - {g^2_P\over2}\int d^4x d^4y
            {\rm Tr}\left[i\gamma_5\Delta(x-y)i\gamma_5\Delta(y-x)
            P(x)P(y)\right]\nonumber\\
            & = & - {3g^2_P\over2\pi^4} \int d^4x d^4y
            {\rm Tr}\left[P(x)P(y)\right]
            {1\over(x-y)^6}\nonumber\\
         &\buildrel sing. \over = & {{3g^2_P}\over{128\pi^4}}\int d^4x d^4y
          \left[{3\over2}P^0(x)P^0(y) + \sum^8_{i=1}
          P^i(x)P^i(y)\right]\nonumber\\
         &   &{\hskip 2cm}\times\Box\Box{\ln(x-y)^2\mu^2\over(x-y)^2}\,,
\end{eqnarray} 
and its logarithmic dependence is
\begin{equation}
{d\Pi_P\over d\ln\mu^2} 
         = -{{3g^2_P}\over{32\pi^2}}\int d^4x
          \left[{3\over2}P^0(x)\Box P^0(x) + \sum^8_{i=1}
          P^i(x)\Box P^i(x)\right]\,.
\end{equation}

The bilinear term in the vector field is
\begin{eqnarray}
\Pi_V(x,y) & = & - {g^2_V\over2}\int d^4x d^4y
       {\rm Tr}\left[i\not\!V(x)\Delta(x-y)i\not\!V(y)
       \Delta(y-x)\right]\nonumber\\
& = & {3g^2_V\over2\pi^4}\int d^4x d^4y
       {\rm Tr}\left[({\lambda^i\over2}\rho^i_\alpha(x)
       + {\lambda^0\over2}\omega_\alpha(x))
       ({\lambda^j\over2}\rho^j_\beta(y)
       + {\lambda^0\over2}\omega_\beta(y))\right]\nonumber\\
&   &\times{(x-y)_\mu(x-y)_\nu\over(x-y)^8}
       {\rm Tr}(\gamma^\alpha\gamma^\mu\gamma^\beta\gamma^\nu)\nonumber\\
&\buildrel sing. \over = & {{3g^2_V}\over{32\pi^4}}\int d^4x d^4y
       \left[{3\over2}\omega^\mu(x)\omega_\mu(y) + \sum^8_{i=1}
       \rho^{i\mu}(x)\rho^i_\mu(y)\right]\nonumber\\
&   &{\hskip 2cm}\times\Box\Box{\ln(x-y)^2\mu^2\over(x-y)^2}\,.
\end{eqnarray}
Its logarithmic dependence is 
\begin{equation}
{d\Pi_V \over d\ln\mu^2} = - {3g^2_V\over8\pi^2}\int d^4x\left[
     {3\over2}\omega^\mu(x)\Box\omega_\mu(x)
     +\sum^8_{i=1}\rho^{i\mu}(x)\Box\rho^i_\mu(x)\right]\,.
\end{equation}

The bilinear term in the axial vector field is
\begin{eqnarray}
\Pi_A(x, y)&\buildrel sing. \over = &{{3g^2_A}\over{32\pi^4}}\int d^4x d^4y
          \left[{3\over2}f^\mu(x)f_\mu(y) + \sum^8_{i=1}
          a^{i\mu}(x)a^i_\mu(y)\right]\nonumber\\
    &  &{\hskip 2cm}\times\Box\Box{\ln(x-y)^2\mu^2\over(x-y)^2}\,,
\end{eqnarray}
and its logarithmic dependence is
\begin{equation}
{d\Pi_A\over d\ln\mu^2} = - {3g^2_A\over8\pi^2}\int d^4x\left[
  {3\over2}f^\mu(x)\Box f_\mu(x) + \sum^8_{i=1}a^{i\mu}(x)\Box a^i_\mu(x)
  \right]\,.
\end{equation}

The one-loop vector-scalar-scalar interaction reads 
\begin{eqnarray}
\Gamma_{VSS}(x, y, z)& = & - g_Vg^2_S\int d^4x d^4y d^4z
     {\rm Tr}\left[-i\not\!V(x)\Delta(x-y)S(y)\right.\nonumber\\
& &  \left.\times\Delta(y-z)S(z)\Delta(z-x)\right]
     + ({\rm permut}.)\nonumber\\
& = & - {3g_Vg^2_S\over32\pi^6}\int d^4x d^4y d^4z
     {\rm Tr}\left[({\lambda^i\over2}\rho^i_\mu(x)
     +{\lambda^0\over2}\omega_\mu(x)){\lambda^j\over2}
     S^j(y)\right.\nonumber\\
& &  \left.\times{\lambda^k\over2}S^k(z)\right]
     {\rm Tr}\left[\gamma^\mu\gamma^\alpha\gamma^\beta\gamma^\gamma\right]
     {(x-y)_\alpha(y-z)_\beta(z-x)_\gamma\over
     (x-y)^4(y-z)^4(z-x)^4}\nonumber\\
&\buildrel sing. \over = & {3g_Vg^2_S\over128\pi^4}\int d^4x d^4y d^4z
     (\omega^\mu(x)\partial_\mu S^i(y)S^i(z)\nonumber\\
& &  +(if_{abc} + d_{abc})\rho^{a\mu}(x)\partial_\mu
       S^b(y)S^c(z))\nonumber\\
& &  \times\left[3\delta(x-z)\Box{\ln(x-y)^2\mu^2\over(x-y)^2}\right.\nonumber\\
& & {\hskip 2cm}\left.+ \delta(x-z)\Box{\ln(y-z)^2\mu^2\over(y-z)^2}\right]\,,
\end{eqnarray}
and the logarithmic dependence is
\begin{eqnarray}
{d\Gamma_{VSS}\over d\ln\mu^2}& = &
      - {g_Vg^2_S\over128\pi^4}\int d^4x d^4y d^4z
      (\omega^\mu(x)\partial_\mu S^i(y)S^i(z)\nonumber\\
& &   + (if_{abc} + d_{abc})\rho^{a\mu}(x)\partial_\mu
       S^b(y)S^c(z))\delta(x-z)\nonumber\\
& &   \times\left[12\pi^2\delta(x-y) + 4\pi^2
      \delta(y-z)\right]\nonumber\\
& = & - {g_Vg^2_S\over8\pi^2}\int d^4 x
      (\omega^\mu(x)\partial_\mu S^i(x)S^i(x)\nonumber\\
& &   + (if_{abc} + d_{abc})\rho^{a\mu}(x)\partial_\mu S^b(x)S^c(x))\,.
\end{eqnarray}

Other one-loop quark contributions are listed in the appendix.

\vskip 1cm
\noindent
{\large \bf 4. Discussion}

We have calculated the one-loop effective action of the ENJL model 
by the method of differential renormalization, which seems to be a very
simple method.  We note that $\gamma_5$ is not well defined in arbitrary 
dimensions of space-time.  But differential renormalization performs 
loop calculation in coordinate space with well defined $\gamma_5$.  
Indeed, the bosonized ENJL model gives the low energy meson effective 
Lagrangian by means of quark loop contributions.  It is a model that 
interpolates between QCD and chiral perturbation theory.  The original 
ENJL Lagrangian which has four-fermion interaction is non-renormalizable,
but the bosonized version is renormalizable.  The anomalous part of the
effective action can be calculated by the WZW method and higher loop can 
be considered further.    

\vskip 1cm
\noindent
{\large \bf Appendix}

Here we list the rest of the one-loop logarithmic quark contributions to the
effective action.
\begin{eqnarray}
{d\Gamma_{VPP}\over d\ln\mu^2} & = &
         {g_Vg^2_S\over8\pi^2}\int d^4x
    \left[\omega^\mu(\partial_\mu P^i)P^i\right.\nonumber\\
 &   & \left.+ (if_{abc} + d_{abc})\rho^{a\mu}(\partial_\mu
         P^b)P^c\right]\,,
\end{eqnarray}
\begin{eqnarray}
\Gamma_{VPS}(x, y, z) = \Gamma_{ASS}(x, y, z) = \Gamma_{APP}(x, y, z) =0\,,
\end{eqnarray}
\begin{eqnarray}
{d\Gamma_{APS}\over d\ln\mu^2} & = &
         {g_Ag_Sg_P\over4\pi^2}\int d^4x
     \left[(\partial^\mu f_\mu)S^iP^i\right.\nonumber\\
 &   & \left.  +(if_{abc} + d_{abc})(\partial^\mu a^a_\mu)
       S^bP^c\right]\,,
\end{eqnarray}
\begin{eqnarray}
{d\Gamma_{AAV}\over d\ln\mu^2}& = &
         {g_Vg^2_A\over8\pi^2}\int d^4x\left\{
       {7\over2}a^{a\mu}(\partial^\nu\rho^a_\nu)f_\mu\right.
     - 4a^{a\mu}\rho^a_\mu\partial^\nu f_\nu\nonumber\\
& &  - {1\over2}a^{a\mu}(\partial^\nu\rho^a_\mu)f_\nu
     - {1\over2}a^{a\mu}(\partial_\mu\rho^{a\nu})f_\nu
     + {7\over2}a^{a\mu}(\partial^\nu\omega_\nu)a^a_\mu\nonumber\\
& &  - 4a^{a\mu}\omega_\mu\partial^\nu a^a_\nu
     - {1\over2}a^{a\mu}(\partial^\nu\omega_\mu)a^a_\nu
     - {1\over2}a^{a\mu}(\partial_\mu\omega_\nu)a^a_\nu\nonumber\\
& &  + {7\over2}f^\mu(\partial^\nu\rho^a_\nu)a^a_\mu
     - 4f^\mu\rho^a_\mu\partial^\nu a^a_\nu
     - {1\over2}f^\mu(\partial^\nu\rho^a_\mu)a^a_\nu\nonumber\\
& &  - {1\over2}f^\mu(\partial_\mu\rho^{a\nu})a^a_\nu
     + {21\over4}f^\mu(\partial^\nu\omega_\nu)f_\mu
     - 6f^\mu\omega_\mu\partial^\nu f_\nu\nonumber\\
& &  - {3\over4}f^\mu(\partial^\nu\omega_\mu)f_\nu
     -{3\over4}f^\mu(\partial_\mu\omega^\nu)f_\nu
     + (if_{abc} + d_{abc})\nonumber\\
& & \times\left({7\over2}a^{a\mu}
      (\partial^\nu\rho^a_\nu)a^a_\mu\right.
     - 4a^{a\mu}\rho^b_\mu\partial^\nu a^c_\nu
     - {1\over2}a^{a\mu}(\partial^\nu\rho^b_\mu)a^c_\nu\nonumber\\
& & \left.\left. - {1\over2}a^{a\mu}(\partial_\mu\rho^{b\nu})
         a^c_\nu\right)\right\}\,,
\end{eqnarray}
\begin{eqnarray}
{d\Gamma_{VVV}\over d\ln\mu^2} &=&
     - {g^3_V\over16\pi^2}\int d^4x\left\{
      \rho^{a\mu}\omega_\mu\partial^\nu\rho^a_\nu\right.
     + {1\over2}\rho^{a\mu}\rho^a_\mu\partial^\nu\omega_\nu\nonumber\\
& &  + \rho^{a\mu}(\partial^\nu\rho^a_\mu)\omega_\nu 
     + \rho^{a\mu}(\partial_\mu\rho^{a\nu})\omega_\nu 
     + \rho^{a\mu}(\partial_\mu\omega^\nu)\rho^a_\nu\nonumber\\
& &  + {3\over4}\omega^\mu\omega_\mu\partial^\nu\omega_\nu
     + {3\over2}\omega^\mu\omega^\nu\partial_\mu\omega_\nu
     - (if_{abc} + d_{abc})\nonumber\\
& &  \times\left({7\over2}\rho^{a\mu}
     (\partial^\nu\rho^b_\nu)\rho^c_\mu\right.
     - 4\rho^{a\mu}\rho^b_\mu\partial^\nu\rho^c_\nu
     - {1\over2}\rho^{a\mu}
      (\partial^\nu\rho^b_\mu)\rho^c_\nu\nonumber\\
& &  \left.\left. - {1\over2}\rho^{a\mu}
      (\partial_\mu\rho^{b\nu})\rho^c_\nu\right)\right\}\,,
\end{eqnarray}
\begin{eqnarray}
{d\Gamma_{SSSS} \over d\ln\mu^2}& = &
         - {g^4_S\over\pi^2}\int d^4x\left\{{1\over12}S^4
    + {1\over16}S^aS^bS^cS^d\right.\nonumber\\
& & \times\left[d_{cde}(if_{abe} + d_{abe}) 
   - d_{bde}(if_{aec} + d_{aec})\right.\nonumber\\
& & \left.\left. +d_{ade}(if_{ebc} + d_{ebc})\right]\right\}\,,
\end{eqnarray}
\begin{equation}
\Gamma_{SSSP} = \Gamma_{SPPP} =0\,,
\end{equation}
\begin{eqnarray}
{d\Gamma_{SSPP}\over d\ln\mu^2} & = &
    - {4g^2_Sg^2_P\over\pi^2}\int d^4x\{{1\over12}
         S^2P^2\nonumber\\
& & + {1\over16}S^aS^bP^cP^d[
      d_{cde}(if_{abe} + d_{abe})\nonumber\\  
& & - d_{bde}(if_{aec} + d_{aec})
    + d_{ade}(if_{ebc} + d_{ebc})]\}\,,
\end{eqnarray}
\begin{eqnarray}
{d\Gamma_{PPPP}\over d\ln\mu^2} & = &
         - {4g^4_P\over\pi^2}\int d^4x\{{1\over12}P^4
    + {1\over16}P^aP^bP^cP^d\nonumber\\
& & \times[d_{cde}(if_{abe} + d_{abe}) 
    - d_{bde}(if_{aec} + d_{aec})\nonumber\\
& & + d_{ade}(if_{ebc} + d_{ebc})]\}\,,
\end{eqnarray}
\begin{eqnarray}
{d\Gamma_{AASS}\over d\ln\mu^2} & = & - {g^2_Sg^2_A\over12\pi^2}
   \int d^4x\left\{{1\over3}a^{a\mu}a^a_\mu S^bS^b
   + {1\over4}a^{a\mu}a^b_\mu S^cS^d
   \left[d_{cde}(if_{abe} + d_{abe})\right.\right.\nonumber\\
&   &\left.- d_{bde}(if_{aec} + d_{dec})
   + d_{ade}(if_{ebc} + d_{ebc})\right]\nonumber\\
&   &\left.+ a^{a\mu}f_\mu S^bS^c
   (if_{abc} + d_{abc})
   + {1\over2}f^\mu f_\mu S^aS^a\right\}\,,
\end{eqnarray}
\begin{eqnarray}
{d\Gamma_{AAPP}\over d\ln\mu^2} & = & - {g^2_Pg^2_A\over12\pi^2}
    \int d^4x\left\{{1\over3}a^{a\mu}a^a_\mu P^bP^b
   + {1\over4}a^{a\mu}a^b_\mu P^cP^d
   \left[d_{cde}(if_{abe} + d_{abe})\right.\right.\nonumber\\
& &\left. - d_{bde}(if_{aec} + d_{dec})
   + d_{ade}(if_{ebc} + d_{ebc})\right]\nonumber\\
& &\left. + a^{a\mu}f_\mu P^bP^c(if_{abc} + d_{abc})
   + {1\over2}f^\mu f_\mu P^aP^a\right\}\,,
\end{eqnarray}
\begin{eqnarray}
{d\Gamma_{VVSS}\over d\ln\mu^2} & = & - {g^2_Sg^2_V
   \over12\pi^2}\int d^4x\left\{{1\over3}\rho^{a\mu}\rho^a_\mu S^bS^b
   + {1\over4}\rho^{a\mu}\rho^b_\mu S^cS^d
   \left[d_{cde}(if_{abe} + d_{abe})\right.\right.\nonumber\\
& &\left.- d_{bde}(if_{aec} + d_{aec})
   + d_{ade}(if_{ebc} + d_{ebc})\right]\nonumber\\
& &\left. + \rho^{a\mu}\omega_\mu S^bS^c(if_{abc} + d_{abc})
   + {1\over2}\omega^\mu\omega_\mu S^aS^a\right\}\,,
\end{eqnarray}
\begin{eqnarray}
{d\Gamma_{VVPP}\over d\ln\mu^2} & = & - {g^2_Pg^2_V\over12\pi^2}
    \int d^4x\left\{{1\over3}\rho^{a\mu}\rho^a_\mu P^bP^b
   + {1\over4}\rho^{a\mu}\rho^b_\mu P^cP^d
   \left[d_{cde}(if_{abe} + d_{abe})\right.\right.\nonumber\\
& &\left. - d_{bde}(if_{aec} + d_{aec})
   + d_{ade}(if_{ebc} + d_{ebc})\right]\nonumber\\
& &\left.+ \rho^{a\mu}\omega_\mu P^b P^c(if_{abc} + d_{abc})
   + {1\over2}\omega^\mu\omega_\mu P^aP^a\right\}\,,
\end{eqnarray}
\begin{eqnarray}
{d\Gamma_{AAVV}\over d\ln\mu^2} &=& {g^2_Ag^2_V\over6\pi^2}\int d^4x
    (2\delta^{\mu\nu}\delta^{\lambda\sigma}
    - 7\delta^{\mu\lambda}\delta^{\nu\sigma}
    - \delta^{\mu\sigma}\delta^{\nu\lambda})\nonumber\\
& & \times\left\{{1\over16}a^a_\mu a^b_\nu\rho^c_\lambda
    \rho^d_\sigma[{4\over3}(\delta^{ab} \delta^{cd}
    - \delta^{bd}\delta^{ac} + \delta^{ad}\delta^{bc})\right.\nonumber\\
& & \left.+ d_{cde}(if_{abe} + d_{abe}) - d_{bde}(if_{aec} + d_{aec})
    + d_{ade}(if_{ebc} + d_{ebc})\right]\nonumber\\
& & + {1\over8}(a^a_\mu a^b_\nu\rho^c_\lambda\omega_\sigma
    + a^a_\mu a^b_\nu\rho^c_\lambda\omega_\sigma
    + a^a_\mu f_\nu\rho^b_\lambda\rho^c_\sigma
    + f_\mu a^a_\nu \rho^b_\lambda\rho^c_\sigma)\nonumber\\
& & \times(if_{abc} + d_{abc}) 
    + {1\over8}(a^a_\mu a^a_\nu\omega_\lambda\omega_\sigma
    + a^a_\mu f_\nu\rho^a_\lambda\omega_\sigma
    + a^a_\mu f_\nu\omega_\lambda\rho^a_\sigma\nonumber\\
& & + f_\mu a^a_\nu\rho^a_\lambda\omega_\sigma
    + f_\mu a^a_\nu\omega_\lambda\rho^a_\sigma
    + f_\mu f_\nu\rho^a_\lambda\rho^a_\sigma)
    \left. + {3\over16}f_\mu f_\nu\omega_\lambda\omega_\sigma
    \right\}\,,
\end{eqnarray}
\begin{eqnarray}
{d\Gamma_{AAAA}\over d\ln\mu^2} &=& {g^4_A\over24\pi^2}\int d^4x
    (2\delta^{\mu\nu}\delta^{\lambda\sigma}
    - 7\delta^{\mu\lambda}\delta^{\nu\sigma}
    - \delta^{\mu\sigma}\delta^{\nu\lambda})\nonumber\\
& & \times\left\{{1\over16}a^a_\mu a^b_\nu a^c_\lambda
    a^d_\sigma[{4\over 3}(\delta^{ab} \delta^{cd}
    - \delta^{bd}\delta^{ac} + \delta^{ad}\delta^{bc})\right.\nonumber\\
& & + d_{cde}(if_{abe} + d_{abe}) 
    - d_{bde}(if_{aec} + d_{aec}) + d_{ade}(if_{ebc} + d_{ebc})]\nonumber\\
& & + {1\over8}(a^a_\mu a^b_\nu a^c_\lambda f_\sigma
    + a^a_\mu a^b_\nu a^c_\sigma f_\lambda
    + a^a_\mu f_\nu a^b_\lambda a^c_\sigma
    + f_\mu a^a_\nu a^b_\lambda a^c_\sigma)\nonumber\\
& & \times(if_{abc} + d_{abc})
    + {1\over8}(a^a_\mu a^a_\nu f_\lambda f_\sigma
    + a^a_\mu f_\nu a^a_\lambda f_\sigma\nonumber\\
& & + a^a_\mu f_\nu f_\lambda a^a_\sigma
    + f_\mu a^a_\nu a^a_\lambda f_\sigma
    + f_\mu a^a_\nu f_\lambda a^a_\sigma
    + f_\mu f_\nu a^a_\lambda a^a_\sigma)\nonumber\\
& & \left. + {3\over16}f_\mu f_\nu f_\lambda f_\sigma\right\}\,,
\end{eqnarray}
\begin{eqnarray}
{d\Gamma_{VVVV}\over d\ln\mu^2} &=& {g^4_V
   \over 24\pi^2}\int d^4 x(2\delta^{\mu\nu}\delta^{\lambda\sigma}
    - 7\delta^{\mu\lambda}\delta^{\nu\sigma}
    - \delta^{\mu\sigma}\delta^{\nu\lambda})\nonumber\\
& & \times\left\{{1\over16}\rho^a_\mu\rho^b_\nu\rho^c_\lambda
    \rho^d_\sigma[{4\over3}(\delta^{ab}\delta^{cd}
    - \delta^{bd}\delta^{ac} + \delta^{ad}\delta^{bc})\right.\nonumber\\
& & + d_{cde}(if_{abe} + d_{abe}) - d_{bde}(if_{aec} + d_{aec})\nonumber\\
& & + d_{ade}(if_{ebc} + d_{ebc})] + {1\over8}(\rho^a_\mu\rho^b_\nu
    \rho^c_\lambda\omega_\sigma + \rho^a_\mu\rho^b_\nu
    \rho^c_\sigma\omega_\lambda\nonumber\\
& & + \rho^a_\mu\omega_\nu\rho^b_\lambda\rho^c_\sigma
    + \omega_\mu\rho^a_\nu\rho^b_\lambda \rho^c_\sigma)
    (if_{abc} + d_{abc})\nonumber\\
& & +{1\over8}(\rho^a_\mu\rho^a_\nu
    \omega_\lambda\omega_\sigma + \rho^a_\mu\omega_\nu
    \rho^a_\lambda\omega_\sigma + \rho^a_\mu\omega_\nu
    \omega_\lambda\rho^a_\sigma + \omega_\mu\rho^a_\nu
    \rho^a_\lambda\omega_\sigma)\nonumber\\
& & + \omega_\mu\rho^a_\nu\omega_\lambda\rho^a_\sigma
    + \omega_\mu\omega_\nu\rho^a_\lambda\rho^a_\sigma)
    \left. + {3\over16}\omega_\mu\omega_\nu
    \omega_\lambda\omega_\sigma\right\}\,.
\end{eqnarray}

\noindent{\large \bf Acknowledgments}

\noindent
This research was supported by the National Science Council of the Republic
of China under Contract Nos. NSC 85-2112-M-006-003 
and NSC 85-2112-M-006-004.

\end{document}